\documentclass[letterpaper, preprint, paper,11pt]{AAS}	

\usepackage{bm}
\usepackage{array}
\usepackage{amsmath}
\usepackage{amssymb}
\usepackage{subfigure}
\usepackage[colorlinks=true, pdfstartview=FitV, linkcolor=black, citecolor= black, urlcolor= black]{hyperref}
\usepackage{overcite}
\usepackage{footnpag}			      	
\usepackage{float}

\PaperNumber{19-293}

\begin{document}

\title{Adaptive Guidance with Reinforcement Meta-Learning}

\author{Brian Gaudet\thanks{Co-Founder, DeepAnalytX LLC, 1130 Swall Meadows Rd, Bishop CA 93514}\hspace{0.1cm} and  
Richard Linares\thanks{Charles Stark Draper Assistant Professor, Department of Aeronautics and Astronautics, Massachusetts Institute of Technology, Cambridge, MA 02139}
}

\maketitle{}

\begin{abstract}
This paper proposes a novel adaptive guidance system developed using reinforcement meta-learning with a recurrent policy and value function approximator. The use of recurrent network layers allows the deployed policy to adapt real time to environmental forces acting on the agent.  We compare the performance of the DR/DV guidance law, an RL agent with a non-recurrent policy, and an RL agent with a recurrent policy in four difficult tasks with unknown but highly variable dynamics. These tasks include a safe Mars landing with random engine failure and a landing on an asteroid with unknown environmental dynamics. We also demonstrate the ability of a recurrent policy to navigate using only Doppler radar altimeter returns, thus integrating guidance and navigation.
\end{abstract}

\section{Introduction}

Many space missions take place in environments with complex and time-varying dynamics that may be  incompletely modeled during the mission design phase. For example, during an orbital refueling mission, the inertia tensor of each of the two spacecraft will change significantly as fuel is transferred from one spacecraft to the other, which can make the combined system difficult to control\cite{guang2018attitude}. The wet mass of an exoatmospheric kill vehicles (EKV) consists largely of fuel, and as this is depleted with divert thrusts, the center of mass changes, and the divert thrusts are no longer orthogonal to the EKV's velocity vector, which wastes fuel and impacts performance. Future missions to asteroids might be undertaken before the asteroid's gravitational field, rotational velocity, and local solar radiation pressure are accurately modeled. Also consider that the aerodynamics of hypersonic re-entry cannot be perfectly modeled. Moreover, there is the  problem of sensor distortion creating bias to the state estimate given to a guidance and control system. Finally, there is the possibility of actuator failure, which significantly modifies the dynamic system of a spacecraft and its environment. These examples show a clear need for a guidance system that can adapt in real time to time-varying system dynamics that are likely to be imperfectly modeled prior to the mission. 

Recently, several works have demonstrated improved performance with uncertain and complex dynamics by training with randomized system parameters. In (Reference~\citenum{yu2017preparing}), the authors use a recurrent neural network to explicitly learn model parameters through real time interaction with an environment; these parameters are then used to augment the observation for a standard reinforcement learning algorithm.  In (Reference~\citenum{peng2017sim}), the authors use a recurrent policy and value function in a modified deep deterministic policy gradient algorithm to learn a policy for a robotic manipulator arm that uses real camera images as observations. In both cases, the agents train over a wide range of randomized system parameters.  In the deployed policy, the recurrent network's internal state quickly adapts to the actual system dynamics, providing good  performance for the agent's respective tasks.

An RL agent using a recurrent network to learn a policy that performs well over a wide range of environmental conditions can be considered a form of meta-learning. In meta-learning (learning to learn), an agent learns through experience on a wide range of Markov decision processes (MDP) a strategy for quickly learning (adapting) to novel MDPs. By considering environments with different dynamical systems as different MDPs, we see that our approach is a form of meta-learning. Specifically, if the system is trained on randomized dynamical systems (as in References \citenum{peng2017sim} and \citenum{yu2017preparing}), this gives the policy experience on a range of ground truth dynamics, and minimizing the cost function requires that the recurrent policy's hidden state quickly adapts to different dynamical systems.

In this work, we use reinforcement meta-learning to develop an adaptive guidance law. Specifically, we will use proximal policy optimization (PPO) with both the policy and value function implementing recurrent layers in their networks.  To understand how recurrent layers result in an adaptive agent, consider that given some agent position and velocity $\mathbf{x}\in\mathbb{R}^6$, and action vector $\mathbf{u}\in\mathbb{R}^3$ output by the agent's policy, the next observation depends not only on $\mathbf{x}$ and $\mathbf{u}$, but also on the ground truth agent mass and external forces acting on the agent. Consequently, during training, the hidden state of a network's recurrent network evolves differently depending on the observed sequence of observations from the environment and actions output by the policy. Specifically, since the policy's weights (including those in the recurrent layers) are optimized to maximize the likelihood of actions that lead to high advantages, the trained policy's hidden state captures unobserved information such as external forces and the current lander mass, as well as the past history of observations and actions, as this information is useful in minimizing the cost function. In contrast, a non-recurrent policy (which we will refer to as an MLP policy), which does not maintain a persistent state vector, can only optimize using a set of current observations, actions, and advantages, and will tend to under-perform a recurrent policy on tasks with randomized dynamics, although as we have shown in (Reference \citenum{gaudet2018deep}), training with parameter uncertainty can give good results using an MLP policy, provided the parameter uncertainty is not too extreme. After training, although the recurrent policy's network weights are frozen, the hidden state will continue to evolve in response to a sequence of observations and actions, thus making the policy adaptive.  In contrast, an MLP policy's behavior is fixed by the network parameters at test time.

An agent implementing recurrent networks can potentially learn a policy for a partially observable Markov decision process (POMPD). To see why, again consider a sequence of observations and actions taken in a trajectory, with the observations not satisfying the Markov property, i.e., $p(o_{t}|a, o_{t-1}, o_{t-2}, ... \ne p(o_{t}|a, o_{t-1})$.  Consequently, minimizing the policy's cost function requires the hidden state representation to contain information on the temporal dependencies in a sequence of observations and actions.  This is similar to how a recursive Bayesian filter can learn to predict a spacecraft's full state from a history of position only measurements. However, a recurrent network has the ability to capture much longer temporal dependencies than a Kalman filter.

We compare the performance of the DR/DV guidance law\cite{d1997optimal}, an RL agent with an MLP policy, and an RL agent with a recurrent policy over a range of tasks with unknown but highly variable dynamics. We use the DR/DV guidance law as a performance baseline, and to improve its performance, we give the DR/DV guidance law access to the ground truth gravitational force, as well as the lander mass at the start of an episode. In contrast, the RL agent only has access to observations that are a function of the lander's position and velocity.  Except for the asteroid landing task, these tasks are variations on the Mars Landing task.  These tasks include:

\begin{enumerate}
    \item \underline{Unknown Dynamics (Asteroid Landing):}  In each episode, the acceleration due to gravity, solar radiation pressure, and rotation are randomly chosen over a wide range, limited only by the lander's thrust capability.
    \item \underline{Engine Failure:}  We assume a redundant thruster configuration (such as used on MSL), and at the start of each episode, with probability $p$ a random thruster is disabled which results in a reduction in thrust capability.
    \item \underline{Large Mass Variation:} We use a large engine specific impulse and assume wet/dry masses of 2000kg/200kg respectively, which results in a large variation in lander mass during the landing. This creates a difficult control problem, as the agent does not have access to the ground truth mass.  
    \item \underline{Landing using non-Markov observations:}  The agent must learn to land using only the returns from the four Doppler radar altimeters as an observation, whereas the previous tasks assumed access to observations that are a function of the ground truth position and velocity. The measurement error becomes quite extreme at lower elevations, making this a difficult task. For obvious reasons, we don't try this with the DR/DV guidance law.
\end{enumerate}

\section{Methods}

\subsection{Equations of Motion}

We model the landing in 3-DOF, where the translational motion is modeled as follows:
\begin{subequations}
\begin{align}
	{\Dot{\mathbf r}} &= {{\mathbf v}}\label{eq:EQOMa}\\
	{\Dot{\bf v}} &= {\frac{{{\bf T}}+{\bf F}_\text{env}}{m} + {\bf g}}\label{eq:EQOMb} + 2\mathbf{\dot{r}_{a}}\times\bm{\omega} + (\bm{\omega}\times\mathbf{r_{a}})\times\bm{\omega} \\
	\Dot{m} &= -\frac{\lVert{{\bf T}}\rVert}{I_\text{sp}g_\text{ref}} \label{eq:EQOMc}
\end{align}
\end{subequations}
Here $\mathbf{r}$ is the lander's position in the target centered reference frame, $\mathbf{r_{a}}$ is the lander's position in the planet (or asteroid) centered reference frame, ${{\bf T}}$ is the lander's thrust vector  $g_\text{ref}=9.8$ $\text{m}/\text{s}^{2}$,  ${\bf g}=\begin{bmatrix} 0 & 0 & -3.7114\end{bmatrix} \text{m}/\text{s}^2$ is used for Mars,  $I_\text{sp}=225$ s, and the spacecraft's mass is $m$.  ${\bf F}_\text{env}$ is a vector of normally distributed random variables representing environmental disturbances such as wind and variations in atmospheric density. $\bm{\omega}$ is a vector of rotational velocities in the planet (or asteroid) centered reference frame, and is set to zero for the Mars landing experiments.  For the Mars landing environment, the minimum and maximum thrust is constrained to be in the range [2000N, 15000N], and the lander's nominal wet mass is 2000kg. For the asteroid landing environment, we assume pulsed thrusters with a thrust capability of 2N along each axis in the target centered reference frame.

\subsection{Reinforcement Learning Overview}

In the RL framework, an agent learns through episodic interaction with an environment how to successfully complete a task by learning a \textit{policy} that maps observations to actions. The environment initializes an episode by randomly generating an internal state, mapping this internal state to an observation, and passing the observation to the agent. These observations could be a corrupted version of the internal state (to model sensor noise) or could be raw sensor outputs such as Doppler radar altimeter readings, or a multi-channel pixel map from an electro-optical sensor.  At each step of the episode, an observation is generated from the internal state, and given to the agent. The agent uses this observation to generate an action that is sent to the environment; the environment then uses the action and the current state to generate the next state and a scalar reward signal.  The reward and the observation corresponding to the next state are then passed to the agent. The environment can terminate an episode, with the termination signaled to the agent via a done signal. The termination could be due to the agent completing the task or violating a constraint.  

Initially, the agent's actions are random, which allows the agent to explore the state space and begin learning the value of experiencing a given observation, and which actions are to be preferred as a function of this observation. Here the value of an observation is the expected sum of discounted rewards received after experiencing that observation; this is similar to the cost-to-go in optimal control. As the agent gains experience, the amount of exploration is decreased, allowing the agent to exploit this experience. For most applications (unless a stochastic policy is required), when the policy is deployed in the field, exploration is turned off, as exploration gets quite expensive using an actual lander.  The safe method of continuous learning in the field is to have the lander send back telemetry data, which can be used to improve the environment's dynamics model, and update the policy via simulated experience.
 
 In the following discussion, the vector ${\bf x}_k$ denotes the observation provided by the environment to the agent. Note that in general ${\bf x}_k$ does not need to satisfy the Markov property. In those cases where it does not, several techniques have proven successful in practice. In one approach, observations spanning multiple time steps  are concatenated, allowing the agent access to a short history of observations, which helps the agent infer the motion of objects in consecutive observations.  This was the approach used in [\citenum{mnih2015human}]. In another approach, a recurrent neural network is used for the policy and value function implementations.  The recurrent network allows the agent to infer motion from observations, similar to the way a recursive Bayesian filter can infer velocity from a history of position measurements. The use of recurrent network layers has proven effective in supervised learning tasks where a video stream needs to be mapped to a label [\citenum{baccouche2011sequential}].
 
 Each episode results in a trajectory defined by observation, actions, and rewards; a step in the trajectory at time $t_k$ can be represented as $({\bf x}_{k},{\bf u}_{k},r_{k})$, where ${\bf x}_k$ is the observation provided by the environment, ${\bf u}_k$ the action taken by the agent using the observation, and $r_k$ the reward returned by the environment to the agent.  The reward can be a function of both the observation ${\bf x}_{k}$ and the action ${\bf u}_k$. The reward is typically discounted to allow for infinite horizons and to facilitate temporal credit assignment. Then the sum of discounted rewards for a trajectory can be defined as
  \begin{equation}\label{objective_RL}
 r({\bm \tau})=\sum_{i=0}^{T}\gamma^i r_k({\bf x}_{k},{\bf u}_k)
  \end{equation}
  where ${\bm \tau}=[{\bf x}_{0},{\bf u}_0,...,{\bf x}_{T},{\bf u}_T]$ denotes the trajectory and $\gamma \in [0,1)$ denotes the discount factor.
 The objective function the RL methods seek to optimize is given by 
 \begin{equation}\label{objective_RL}
 J({\bm \theta})=\mathbb{E}_{p({\bm \tau})}\left[ r({\bm \tau})\right]=\int_{\mathbb{T}}r({\bm \tau})p_{\bm \theta}(\tau)d{\bm \tau}
 \end{equation}
where 
\begin{equation}
p_{\bm \theta}(\bm\tau)=\left[ \prod_{k=0}^{T}p({\bf x}_{k+1}|{\bf x}_k,{\bf u}_{\bm \theta})\right]p({\bf x}_0)
 \end{equation}
 where $\mathbb{E}_{p({\bm \tau})}\left[\cdot\right]$ denotes the expectation over trajectories and in general ${\bf u}_{\bm \theta}$ may be deterministic or stochastic function of the policy parameters, ${\bm \theta}$. However, it was noticed by Ref. \citenum{williams1992simple} that if the policy is chosen to be stochastic, where ${\bf u}_k\sim {\pi}_{\bm \theta}({\bf u}_k|{\bf x}_k)$ is a pdf for ${\bf u}_k$ conditioned on ${\bf x}_k$, then a simple policy gradient expression can be found. 
  \begin{equation}\label{policy_gradient_reinforce}
  \begin{aligned}
 \nabla_{\bm \theta} J({\bm \theta})=&\int_{\mathbb{T}} \sum_{k=0}^{T}r_k({\bf x}_k,{\bf u}_k)  \nabla_{\bm \theta} \log  {\pi}_{\bm \theta}({\bf u}_k|{\bf x}_k) p_{\bm \theta}({ \bm \tau})d{\bm \tau}\\
 &\approx \sum_{i=0}^{M}\sum_{k=0}^{T}r_k({\bf x}_k^i,{\bf u}_k^i)  \nabla_{\bm \theta} \log  {\pi}_{\bm \theta}({\bf u}_k^i|{\bf x}_k^i)
 \end{aligned}
 \end{equation}
 where the integral over ${\bm \tau}$ is approximated with samples from ${\bm \tau}^i\sim p_{\bm \theta}({\bm \tau})$ which are monte carlo roll-outs of the policy given the environment's transition pdf, $p({\bf x}_{k+1}|{\bf x}_k)$. The expression in Eq.~\eqref{policy_gradient_reinforce} is called the policy gradient and the form of this equation is referred to as the REINFORCE method [\citenum{williams1992simple}]. Since the development of the REINFORCE method additional theoretical work improved on the performance of the REINFORCE method. In particular, it was shown that the reward $r_k({\bf x}_k,{\bf u}_k)$ in Eq.~\eqref{policy_gradient_reinforce}  can be replaced with state-action value function $Q^{\pi}({\bf x}_k,{\bf u}_k)$, this result is known as the Policy Gradient Theorem. Furthermore, the variance of the policy gradient estimate that is derived from the monte carlo roll-outs, ${\bm \tau}^i$, is reduced by subtracting a state-dependent basis from $Q^{\pi}({\bf x}_k,{\bf u}_k)$. This basis is commonly chosen to be the state value function $V^{\pi}({\bf x}_k)$, and we can define $A^{\pi}({\bf x}_k,{\bf u}_k)=Q^{\pi}({\bf x}_k,{\bf u}_k)-V^{\pi}({\bf x}_k)$ . This method is known as the Advantage-Actor-Critic (A2C) Method. The policy gradient for the A2C method is given by (where the  $\mathbf w$ subscript denotes a function parameterized by $\mathbf w$)
 \begin{equation}
\label{eq:pggrad}
\nabla_{{\bm \theta}}J({\bm \theta})\approx \sum_{i=0}^{M}\sum_{k=0}^{T}A^{\pi}_{\bf{w}}({\bf x}_{k}^i,{\bf u}_{k}^i)  \nabla_{\bm \theta} \log  {\pi}_{\bm \theta}({\bf u}_k^i|{\bf x}_k^i)
\end{equation}

\subsection{Proximal Policy Optimization} \label{PGRL}
The Proximal Policy Optimization (PPO) approach  [\citenum{schulman2017proximal}] is a type of policy gradient which has demonstrated state-of-the-art performance for many RL benchmark problem. The PPO approach is developed using the properties of the Trust Region Policy Optimization (TRPO) Method [\citenum{schulman2015trust}]. The TRPO method formulates the policy optimization problem using a constraint to restrict the size of the gradient step taken during each iteration [\citenum{sorensen1982newton}]. The TRPO method policy update is calculated using the following problem statement:
\begin{equation}\label{TRPOeq}
\begin{aligned}
& \underset{{\bm \theta}}{\text{minimize}}
& & \mathbb{E}_{p({\bm \tau})}\left[\frac{\pi_{{\bm \theta}}({\bf u}_{k}|{\bf x}_{k})}{\pi_{{\bm \theta}_\text{old}}({\bf u}_{k}|{\bf x}_{k})}A^{\pi}_{\bf w}({\bf x}_{k},{\bf u}_{k})\right] \\
& \text{subject to}
& & \mathbb{E}_{p({\bm \tau})}\left[ \text{KL}\left( \pi_{{\bm \theta}}({\bf u}_{k}|{\bf x}_{k}),\pi_{{\bm \theta}_\text{old}}({\bf u}_{k}|{\bf x}_{k}) \right)  \right] \leqslant \delta
\end{aligned}
\end{equation}
The parameter $\delta$ is a tuning parameter but the theory justifying the TRPO methods proves monotonic improvement in the policy performance if the policy change in each iteration is bounded a parameter $C$. The parameter $C$ is computed using the Kullback-Leibler (KL) divergence [\citenum{kullback1951information}]. Reference \citenum{schulman2015trust} computes a closed-form expression for $C$ but this expression leads to prohibitively small steps, and therefore, Eq.~\eqref{TRPOeq} with a fix constraint is used. Additionally, Eq.~\eqref{TRPOeq} is approximately solved using the conjugate gradient algorithm, which approximates the constrained optimization problem given by Eq.~\eqref{TRPOeq} with a linearized objective function and a quadratic approximation for the constraint. The PPO method approximates the TRPO optimization process by accounting for the policy adjustment constrain with a clipped objective function. The objective function used with PPO can be expressed in terms of the probability ratio $p_{k}({\bm \theta})$ given by,
\begin{equation}
\label{eq:clipr}
p_{k}({\bm \theta})=\frac{\pi_{{\bm \theta}}({\bf u}_{k}|{\bf x}_{k})}{\pi_{{\bm \theta}_\text{old}}({\bf u}_{k}|{\bf x}_{k})}
\end{equation}
where the PPO objective function is then as follows:
\begin{equation}
\label{eq:ppoloss}
L({\bm \theta})=\mathbb{E}_{p({\bm \tau})}\left[\mathrm{min}\left[p_{k}({\bm \theta}) , \mathrm{clip}(p_{k}({\bm \theta}) , 1-\epsilon, 1+\epsilon)\right]A^{\pi}_{\bf w}({\bf x}_{k},{\bf u}_{k})\right]
\end{equation}
This clipped objective function has been shown to maintain the KL divergence constraints, which aids convergence by insuring that the policy does not change drastically between updates.

PPO uses an approximation to the advantage function that is the difference between the empirical return and a state value function baseline, as shown below in Equation \eqref{eq:ppo_adv}:

\begin{equation}
\label{eq:ppo_adv}
	A^{\pi}_{\bf w}(\mathbf{x}_{k},\mathbf{u}_{k})=\left[\sum_{\ell=k}^{T}\gamma^{\ell-k}r(\bf x_{\ell},\bf u_{\ell})\right]-V_{\bf w}^{\pi}(\mathbf{x}_{k})
\end{equation}
Here the value function $V_{\bf w}^{\pi}$ is learned using the cost function given below in \eqref{eq:vf_ppo}.
\begin{equation}
\label{eq:vf_ppo}
L(\mathbf{w})=\sum_{i=1}^{M}\left(V_{\mathbf{w}}^{\pi}({\bf x}_k^i)-\left[\sum_{\ell=k}^{T}\gamma^{\ell-k}r({\bf u}_{\ell}^i,{\bf x}_{\ell}^i)\right]\right)^2
\end{equation}
In practice, policy gradient algorithms update the policy using a batch of trajectories (roll-outs) collected by interaction with the environment. Each trajectory is associated with a single episode, with a sample from a trajectory collected at step $k$ consisting of observation ${\bf x}_{k}$, action ${\bf u}_{k}$, and reward $r_k({\bf x}_k,{\bf u}_k)$. Finally, gradient accent is performed on ${\bm \theta}$ and gradient decent on ${\bf w}$ and update equations are given by 
\begin{align}\label{loss}
{\bf w}^+&={\bf w}^--\beta_{{\bf w}}\nabla_{{\bf w}} \left. L({\bf w})\right|_{{\bf w}={\bf w}^-}\\
{\bm \theta}^+&={\bm \theta}^-+\beta_{{\bm \theta}} \left. \nabla_{\bm \theta}J\left({\bm \theta}\right)\right|_{{\bm \theta}={\bm \theta}^-}
\end{align}
where $\beta_{{\bf w}}$ and $\beta_{{\bm \theta}}$ are the learning rates for the value function, $V_{\bf w}^{\pi}$, and policy, $\pi_{\bm \theta}\left({\bf u}_k|{\bf x}_k\right)$, respectively.
 
In our implementation, we adjust the clipping parameter $\epsilon$ to target a KL divergence between policy updates of 0.001. The policy and value function are learned concurrently, as the estimated value of a state is policy dependent. We use a Gaussian distribution with mean $\pi_{\bm \theta}({\bf x}_{k})$ and a diagonal covariance matrix for the action distribution in the policy.  Because the log probabilities are calculated using the exploration variance, the degree of exploration automatically adapts during learning such that the objective function is maximized.

\subsection{Recurrent Policy and Value Function Approximator}

Since the policy's cost function uses advantages (the difference between empirical discounted rewards received after taking an action and following the policy and the expected value of the observation from which the action is taken), we also need the agent to implement a recurrent value function approximator. This recurrent value function has a recurrent layer, where the hidden state evolves differently depending on the past sequence of observations and returns.

During learning, we need to unroll the recurrent layer in time in order to capture the temporal relationship between a sequence of observations and actions.  However, we want to do this unrolling in a manner consistent with processing a large number of episodes in parallel. To see why, imagine we want to unroll the network for 60 steps for the forward pass through the network. The hidden state at step 61 is not available until we have completed the forward pass from steps 1 through 60; consequently if we want to do both segments in parallel, we must insert the state from step 60 that occurred when sampling from the policy during optimization as the initial state prior to the forward pass for steps 60-61. 

It follows that to allow parallel computation of the forward pass, we must capture the actual hidden state when we sample from the policy (or predict the value of an observation using the value function) and add it to the rollouts. So where before a single step of interaction with the environment would add the tuple $(\mathbf{o}, \mathbf{a}, r)$ (observation, action, reward) to the rollouts, we would augment this tuple with the hidden states of the policy and value functions.  The forward pass through the policy and value function networks is then modified so that prior to the the recurrent layer, the  network unrolls the output of the previous layer, reshaping the data from $\mathbb{R}^{m\times n}$ to $\mathbb{R}^{T\times m/T\times n}$, where $m$ is the batch size, $n$ the feature dimension, and $T$ the number of steps we unroll the network. At step zero, we input the hidden state from the rollouts to the recurrent layer, but for all subsequent steps up to $T$, we let the state evolve according the current parameterization of the recurrent layer. After the recurrent layer (or layers), the recurrent layer output is then reshaped back to $\mathbb{R}^{m\times n}$. Note that without injecting the hidden states from the rollouts at the first step, parallel unrolling of a batch would not be possible, which would dramatically increase computation time.  

\begin{figure}[H]
\begin{center}
\includegraphics[width=.9\linewidth]{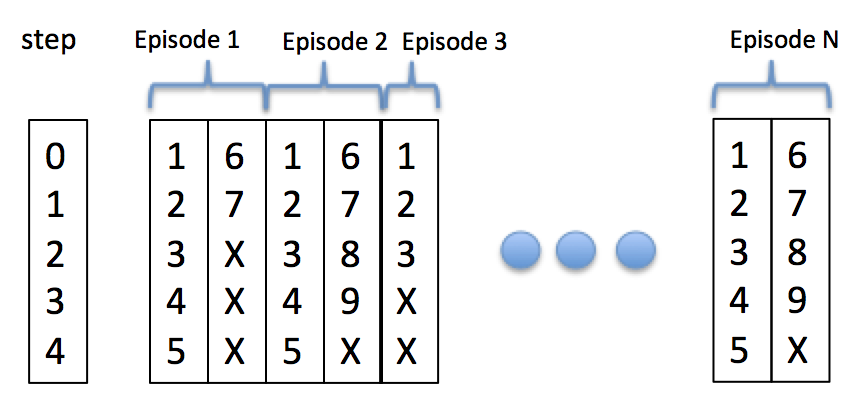}
\caption{Unrolling the forward pass in Time}
\label{fig:unroll}
\end{center}
\end{figure}

There are quite a few implementation details to get this right, as the rollouts must be padded so that $m/T$ is an integer, but unpadded for the computation of the loss. The unrolling in time is illustrated graphically in Figure \ref{fig:unroll} for the case where we unroll for 5 steps. The numbers associated with an episode are the representation of the observation for that step of the trajectory as given by the output of the layer preceding the recurrent layer, and an "X" indicates a padded value. It is also important to not shuffle the training data, as this destroys the temporal association in the trajectories captured in the rollouts. In order to avoid issues with exploding / vanishing gradients when we back propagate through the unrolled recurrent layer, we implement the recurrent layers as gated recurrent units (GRU)\cite{chung2015gated}. 

\subsection{RL Problem Formulation}

A simplified view of the agent and environment are shown below in Figure \ref{fig:a_e_i}.  The policy and value functions are implemented using four layer neural networks with tanh activations on each hidden layer. Layer 2 for the policy and value function is a recurrent layer implemented as a gated recurrent unit. The network architectures are as shown in Table \ref{tab:NN}, where $n_{\mathrm{hi}}$ is the number of units in layer $i$, $\mathrm{obs\_dim}$ is the observation dimension, and $\mathrm{act\_dim}$ is the action dimension.

\begin{figure}[htbp]
\begin{center}
\includegraphics[width=.9\linewidth]{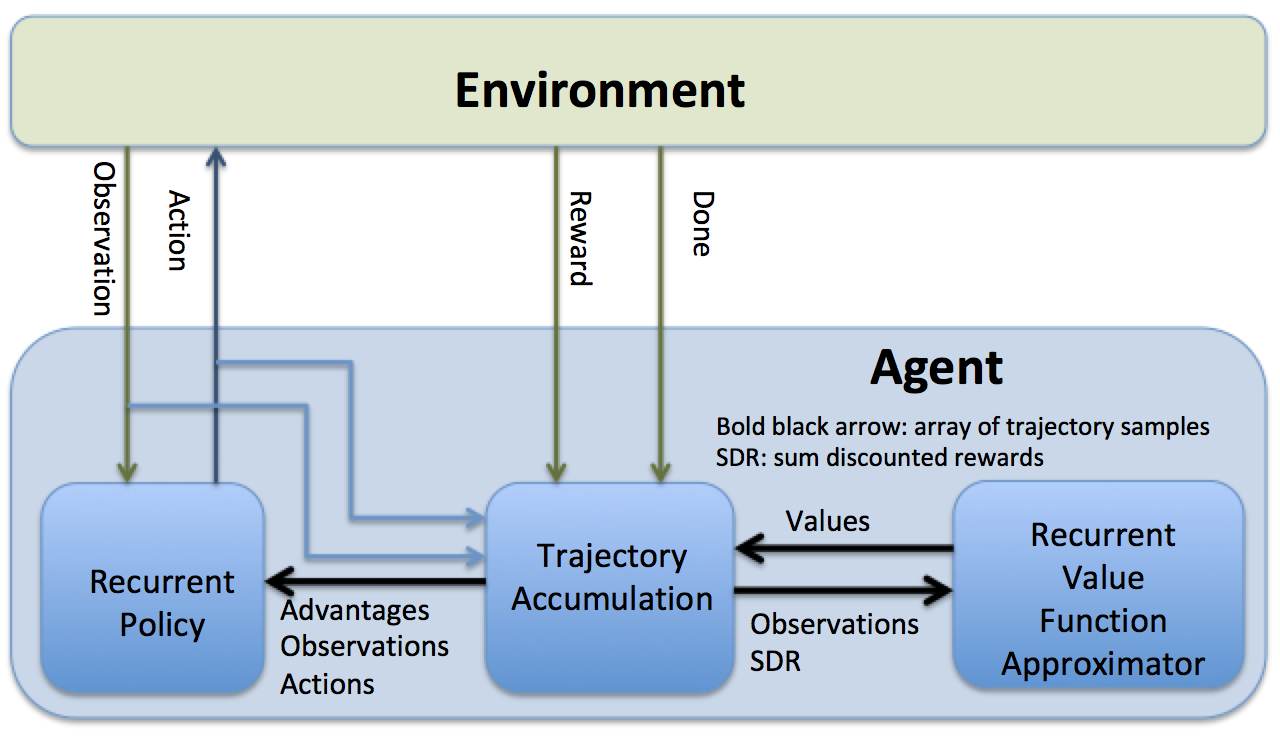}
\caption{Agent-Environment Interface}
\label{fig:a_e_i}
\end{center}
\end{figure}

\begin{table}[htbp]
	\fontsize{10}{10}\selectfont
    \caption{Policy and Value Function network architecture}
   \label{tab:NN}
        \centering 
   \newcolumntype{R}{>{\raggedleft\arraybackslash}p{1.8cm}}
   \begin{tabular}{l | R | c | R | c } 
      \hline 
       & \multicolumn{2}{c}{Policy Network}\vline & \multicolumn{2}{c}{Value Network}\\
       \hline
       Layer & \# units & activation & \# units & activation \\
       \hline
      hidden 1      & $10 * \mathrm{obs\_dim}$ & tanh & $10 * \mathrm{obs\_dim}$ & tanh \\
      hidden 2      & $\sqrt{n_{\mathrm{h1}} * n_{\mathrm{h3}}}$ & tanh & $\sqrt{n_{\mathrm{h1}} * n_{\mathrm{h3}}}$ & tanh\\
      hidden 3      & $10 * \mathrm{act\_dim}$ & tanh & 5 & tanh \\
      output        & $\mathrm{act\_dim}$ & linear & 1 & linear \\
      \hline
   \end{tabular}
\end{table}

The most difficult part of solving the planetary landing problem using RL was the development of a reward function that works well in a sparse reward setting. If we only reward the agent for making a soft pinpoint landing at the correct attitude and with close to zero rotational velocity, the agent would never see the reward within a realistic number of episodes, as the probability of achieving such a landing using random actions in a 3-DOF environment with realistic initial conditions is exceedingly low. The sparse reward problem is typically addressed using inverse reinforcement learning [\citenum{ng2000algorithms}], where a per-timestep reward function is learned from expert demonstrations. With a reward given at each step of agent-environment interaction, the rewards are no longer sparse. 

Instead, we chose a different approach, where we engineer a reward function that, at each time step, provides hints to the agent (referred to as “shaping rewards”) that drive it towards a soft pinpoint landing. The recommended approach for such a reward shaping function is to make the reward a difference of potentials, in which case there are theoretical results showing that the additional reward does not change the optimal policy[\citenum{ng2003shaping}].  We experimented with several potential functions with no success. Instead we drew inspiration from biological systems that use the gaze heuristic. The gaze heuristic is used by animals such as hawks and cheetahs to intercept prey (and baseball players to catch fly balls), and works by keeping the line of sight angle constant during the intercept.  The gaze heuristic is also the basis of the well known PN guidance law used for homing phase missile guidance.  

In our case, the landing site is not maneuvering, and we have the additional constraint that we want the terminal velocity to be small. Therefore we use a heuristic where the agent attempts to keep its velocity vector aligned with the line of sight vector. Since the target is not moving in the target-centered reference frame, the target's future position is its current position, and the optimal action is to head directly towards the target. Such a rule results in a pinpoint, but not necessarily soft, landing.  To achieve the soft landing, the agent estimates time-to-go as the ratio of the range and the magnitude of the lander's velocity, and reduces the targeted velocity as time-to-go decreases.  It is also important that the lander's terminal velocity be directed predominantly downward. To achieve these requirements, we use the piecewise reward shaping function given below in Equations \eqref{eq:vtarg1a}, \eqref{eq:vtarg1b}, \eqref{eq:vtarg1c}, \eqref{eq:vtarg1d}, and \eqref{eq:vtarg1e}, where $\tau_{1}$ and $\tau_{2}$ are hyperparameters  and $v_{o}$ is set to the magnitude of the lander's velocity at the start of the powered descent phase. We see that the shaping rewards take the form of a velocity field that maps the lander's position to a target velocity. In words, we target a location 15m above the desired landing site and target a z-component of landing velocity equal to -2m/s. Below 15m, the downrange and crossrange velocity components of the target velocity field are set to zero, which encourages a vertical descent. 

\begin{subequations}
\begin{align}
    {\bf v}_{targ}&=-v_{o}\left(\frac{{\bf \hat{r}}}{\lVert{\bf \hat{r}}\rVert}\right)\left(1-\exp\left(-\frac{t_{go}}{\tau}\right)\right)\label{eq:vtarg1a}\\
    t_{go}&=\frac{\lVert{\bf \hat{r}}\rVert}{\lVert{\bf \hat{v}}\rVert}\label{eq:vtarg1b} \\
    {\bf \hat{r}}&=\begin{cases}{\bf r} - \begin{bmatrix}0 & 0 & 15\end{bmatrix}, & \text{if } r_{2} > 15 \\ \begin{bmatrix}0 & 0 & r_{2}\end{bmatrix}, &\text{otherwise}\end{cases}\label{eq:vtarg1c} \\
    {\bf \hat{v}}&=\begin{cases}{\bf v} - \begin{bmatrix}0 & 0 & -2\end{bmatrix}, & \text{if } r_{2} > 15 \\ {\bf v}-\begin{bmatrix}0 & 0 & -1\end{bmatrix}, &\text{otherwise}\end{cases}\label{eq:vtarg1d} \\
    \tau&=\begin{cases} \tau_{1}, & \text{if } r_{2} > 15 \\ \tau_{2}, &\text{otherwise}\end{cases}\label{eq:vtarg1e} 
\end{align}
\end{subequations}
Finally, we provide a terminal reward bonus when the lander reaches an altitude of zero, and the terminal position, velocity, and glideslope are within specified limits. The reward function is then given by Equation \eqref{eq:reward_func}, where the various terms are described in the following:

\begin{enumerate}
    \item $\alpha$ weights a term penalizing the error in tracking the target velocity.
    \item $\beta$ weights a term penalizing control effort.
    \item $\gamma$ is a constant positive term that encourages the agent to keep making progress along the trajectory. Since all other rewards are negative, without this term, an agent would be incentivized to violate the attitude constraint and prematurely terminate the episode to maximize the total discounted rewards received starting from the initial state.
    \item $\eta$ is a bonus given for a successful landing, where terminal position, velocity, and glideslope are all within specified limits.  The limits are $r_{lim}=5$ $\text{m}$, $v_{lim}=2$ $\text{m}$,  and $gsmin_{lim}=5$ rad/s. The minimum glideslope at touchdown insures the lander's velocity is directed predominatly downward.
\end{enumerate}
\begin{equation}
 \begin{aligned}
 \label{eq:reward_func}
r &= \alpha\|{\bf v}-{\bf v}_{targ}\|+
\beta\|{\bf T}\|+ 
+\gamma\\
&+\eta({\bf r}_{2}<0 \ \mathrm{and}\  
\|{\bf r}\|<r_{lim} \ \mathrm{and}\  \|{\bf v}\|<v_{lim} \ \mathrm{and}\ 
gs > gsmin_{lim}
\end{aligned}
\end{equation}
This reward function allows the agent to trade off between tracking the target velocity given in Eq.~\eqref{eq:vtarg1a}, conserving fuel, and maximizing the reward bonus given for a good landing.  Note that the constraints are not hard constraints such as might be imposed in an optimal control problem solved using collocation methods. However, the consequences of violating the constraints (a large negative reward and termination of the episode) are sufficient to insure they are not violated once learning has converged. Hyperparameter settings and coefficients used in this work are given below in Table~\ref{tab:HPS}, note that due to lander symmetry, we do not impose any limits on the lander's yaw.
\begin{table}[h]
	\fontsize{10}{10}\selectfont
    \caption{Hyperparameter Settings}
   \label{tab:HPS}
        \centering 
   \begin{tabular}{ c | c | c | c | c | c | c } 
      \hline
      $v_{o}$ (m/s) & $\tau_{1}$ (s) & $\tau_{2}$ (s) & $\alpha$  & $\beta$   &  $\gamma$ & $\eta$\\
      \hline
       $\|{\bf v}_{o}\|$ & 20 & 100 & -0.01    &  -0.05 & 0.01 & 10\\
      
   \end{tabular}
\end{table}
As shown below in Eq.~\eqref{eq:obs}, the observation given to the agent during learning and testing is ${\bf v}_\text{error}={\bf v}-{\bf v}_\text{targ}$, with ${\bf v}_\text{targ}$ given in Eq.~\eqref{eq:vtarg1a} , the lander's estimated altitude, and the time to go.  Note that aside from the altitude, the  lander translational coordinates do not appear in the observation. This results in a policy with good generalization in that the policy's behavior can extend to areas of the full state space that were not experienced during learning.

\begin{equation}
\label{eq:obs}
{\text{obs}}=\begin{bmatrix}{\bf v}_\text{error}  & r_{2} & t_\text{go}\end{bmatrix}
\end{equation}

It turns out that the when a terminal reward is used (as we do), it is advantageous to use a relatively large discount rate.  However, it is also advantageous to use a lower discount rate for the shaping rewards.  To our knowledge, a method of resolving this conflict has not been reported in the literature.  In this work, we resolve the conflict by introducing a framework for accommodating multiple discount rates.  Let $\gamma_{1}$ be the discount rate used to discount $r_{1}(k)$, the reward function term (as given in \ref{eq:reward_func})  associated with the $\kappa$ coefficient). Moreover, let $\gamma_{2}$ be the discount rate used to discount $r_{2}(k)$, the sum of all other terms in the reward function. We can then rewrite the Eq.~\eqref{eq:vf_ppo} and \eqref{eq:ppo_adv} in terms of these rewards and discount rates, as shown below in Eq.~\eqref{eq:new_vf} and Eq.~\eqref{eq:new_adv}.  Although the approach is simple, the performance improvement is significant, exceeding that of using generalized advantage estimation (GAE) [\citenum{schulman2015high}] both with and without multiple discount rates.  Without the use of multiple discount rates, the performance was actually worsened by including the terminal reward term. 
\begin{subequations}
\begin{align}
J({\mathbf{w}})&=\sum_{i=1}^M \left(V_{\bm \theta}^{\pi}({\bf x}_{k})-\left[\sum_{\tau=t}^{n}\gamma_{1}^{\tau-t}r_{1}({\bf u}_{\tau},{\bf x}_{\tau})+\gamma_{2}^{\tau-t}r_{2}({\bf u}_{\tau},{\bf x}_{\tau})\right]\right)^2\label{eq:new_vf}\\
A^{\pi}_{\mathbf{w}}({\bf x}_{t},{\bf u}_{t})&=\left[\sum_{\tau=t}^{n}\gamma_{1}^{\tau-t}r_{1}({\bf u}_{\tau},{\bf x}_{\tau})+\gamma_{2}^{\tau-t}r_{2}({\bf u}_{\tau},{\bf x}_{\tau})\right]-V_{\mathbf{w}}^{\pi}({\bf x}_{k})\label{eq:new_adv}
\end{align}
\end{subequations}

\section{Experiments}

For each experiment, we compare the performance of a DR/DV policy, MLP policy, and recurrent policy with unrolling the recurrent layer through 1, 20, and 60 timesteps during the forward pass through the policy network. As a shorthand, we will refer to a recurrent network with the forward pass unrolled T  steps in time as a T-step RNN. The DR/DV policy just instantiates a DR/DV controller. For the experiments that use the Mars landing environment, each episode begins with the initial conditions shown in Table \ref{tab:IC}. In each experiment, we optimize the policy using PPO, and then test the trained policy for 10,000 episodes. For the experiments using the Mars landing environment, the acceleration due to gravity (including downrange and crossrange components) is randomly set over a uniform distribution +/-5\% of nominal, and the lander's initial mass is set over  a random distribution +/-10\% of nominal.

\begin{table}[H]
	\fontsize{10}{10}\selectfont
    \caption{Mars Lander Initial Conditions for Optimization}
   \label{tab:IC}
        \centering 
   \newcolumntype{R}{>{\raggedleft\arraybackslash}p{1.8cm}}
   \begin{tabular}{l | R | R | R | R } 
      \hline 
       & \multicolumn{2}{c}{Velocity}\vline & \multicolumn{2}{c}{Position}\\
       \hline
       & min (m/s) & max (m/s) & min (m) & max (m) \\
       \hline
      Downrange      & -70 & -10 & 0 & 2000\\
      Crossrange       & -30  & 30 & -1000 & 1000 \\
      Elevation     & -90 & -70 & 2300 & 2400 \\
   \end{tabular}
\end{table}

\subsection{Experiment 1: Asteroid Landing with Unknown Dynamics}

We demonstrate the adaptive guidance system in a simulated landing on an asteroid with unknown and highly variable environmental dynamics. We chose an asteroid landing environment for this task because the asteroid's rotation can cause the Coriolis and centrifugal forces to vary widely, creating in effect an unknown dynamical environment, where the forces are only bounded by the lander's thrust capability (recurrent policies are great, but they can't get around the laws of physics). At the start of each episode, the asteroid's angular velocity ($\omega$), gravity ($\mathbf{g}$), and the local solar radiation pressure (SRP) are randomly chosen within the bounds given in Table \ref{tab:exp1_forces}. Note that  $\omega$, $\mathbf{g}$, and SRP are drawn from a uniform distribution with independent directional components. Near-earth asteroids would normally have parameters known much more tightly than this, but the unknown dynamics scenario could be realistic for a mission that visits many asteroids, and we want the guidance law to adapt to each unique environment. The lander has pulsed thrusters with a 2N thrust capability per direction, and a wet mass that is randomly chosen at the start of each episode within the range of 450kg to 500kg.

\begin{table}[htbp]
	\fontsize{10}{10}\selectfont
    \caption{Experiment 1: Environmental Forces and Lander Mass}
   \label{tab:exp1_forces}
        \centering 
   \newcolumntype{R}{>{\raggedleft\arraybackslash}p{1.5cm}}
   \begin{tabular}{l | R | R | R | R | R | R } 
      \hline 
       & \multicolumn{3}{c}{Minimum}\vline & \multicolumn{3}{c}{Maximum}\\
       \hline
       & $x$ & $y$ & $z$ & $x$ & $y$ & $z$ \\
       \hline
       Asteroid $\omega$ (rad/s) & -1e-3 & -1e-3 & -1e-3 & 1e-3  & 1e-3 & 1e-3 \\
       Asteroid $\mathbf{g}$ $ (m/s^{2}$ &  -1e-6 & -1e-6 & -1e-6 & -100e-6 & -100e-6 & -100e-6  \\
       SRP ($m/s^{2}$)  & -1e-6  &  -1e-6 & -1e-6 & 1e-6 & 1e-6 & 1e-6 \\
   \end{tabular}
\end{table}

The lander's initial conditions are shown in Table \ref{tab:exp1_ic}. The variation in initial conditions is typically much less than this for such a mission.  Indeed, (Reference \citenum{udrea2012sensitivity}) assume position and velocity standard deviations at the start of the Osiris Rex TAG maneuver of 1cm and 0.1cm/s respectively.  The lander targets a position on the asteroid's pole that is a distance of 250m from the asteroid center. Due to the range of environmental parameters tested, the effect would be identical if the target position was on the equator, or anywhere else 250m from the asteroid's center of rotation. For purposes of computing the Coriolis and centrifugal forces, we translate the lander's position from the target centered reference frame to the asteroid centered reference frame.  We define a landing plane with an surface normal depending on the targeted landing site, which allows use of the Mars landing environment with minimal changes.

\begin{table}[htbp]
	\fontsize{10}{10}\selectfont
    \caption{Experiment 1: Lander Initial Conditions}
   \label{tab:exp1_ic}
        \centering 
   \newcolumntype{R}{>{\raggedleft\arraybackslash}p{1.8cm}}
   \begin{tabular}{l | R | R | R | R } 
      \hline 
       & \multicolumn{2}{c}{Velocity}\vline & \multicolumn{2}{c}{Position}\\
       \hline
       & min (cm/s) & max (cm/s) & min (m) & max (m) \\
       \hline
      Downrange      & -100 & -100 & 900 & 1100\\
      Crossrange     & -100 & -100 & 900 & 1100 \\
      Elevation     & -100 & -100 & 900 & 1100 \\
   \end{tabular}
\end{table}

The RL implementation is similar to that of the Mars landing. We used the reward shaping function as shown below in Equations \eqref{eq:ast_vtarg1a} and \eqref{eq:ast_vtarg1b}, the reward function as given in \eqref{eq:reward_func} but with  $gsmin$ set to 0, and the hyperparameter settings shown below in Table \ref{tab:ast_HPS}.  For the terminal reward, we use $r_{lim}=1 m$ and $v_{lim}=0.2 m/s$.  Pulsed thrust was achieved by discretizing the policy output in each dimension to take three values: negative  maximum thrust, zero thrust, or positive maximum thrust. We tuned the nominal gravity parameter for the DR/DV policy for best performance; it turns out that the optimal setting for this parameter is quite a bit higher than the actual environmental gravity.

\begin{subequations}
\begin{align}
    {\bf v}_{targ}&=-v_{o}\left(\frac{{\bf \hat{r}}}{\lVert{\bf \hat{r}}\rVert}\right)\left(1-\exp\left(-\frac{t_{go}}{\tau}\right)\right)\label{eq:ast_vtarg1a}\\
    t_{go}&=\frac{\lVert{\bf \hat{r}}\rVert}{\lVert{\bf \hat{v}}\rVert}\label{eq:ast_vtarg1b} 
\end{align}
\end{subequations}

\begin{table}[htbp]
	\fontsize{10}{10}\selectfont
    \caption{Asteroid Mission Hyperparameter Settings}
   \label{tab:ast_HPS}
        \centering 
   \begin{tabular}{ c | c | c | c | c | c  } 
      \hline
      $v_{o}$ (m/s) & $\tau$ (s)  & $\alpha$  & $\beta$   &  $\gamma$ & $\eta$\\
      \hline
       1 & 300 & -1.0    &  -0.01 & 0.01 & 10\\
      
   \end{tabular}
\end{table}

\begin{figure}[htbp]
\begin{center}
\includegraphics[width=.9\linewidth]{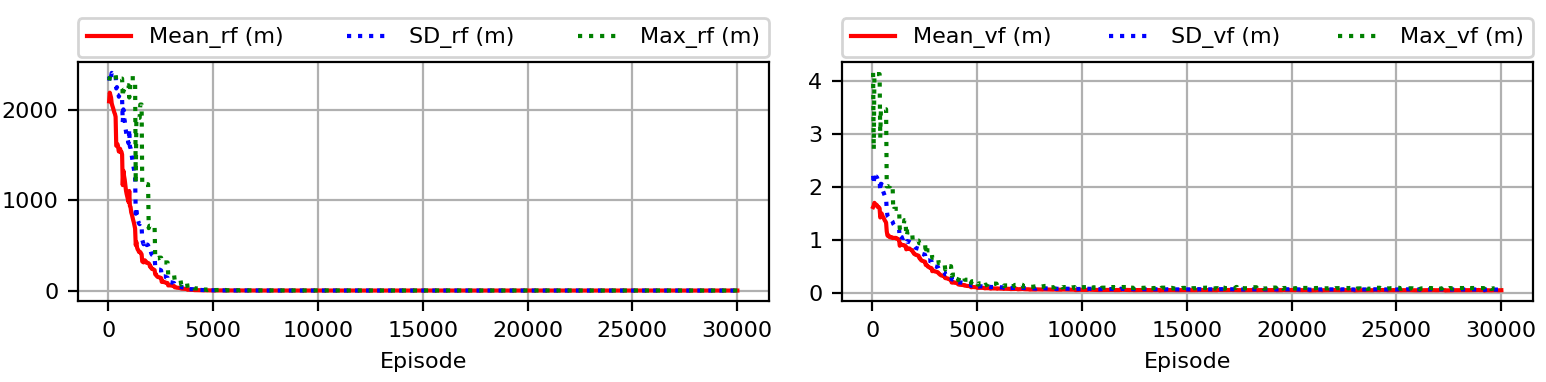}
\caption{Experiment 1: Learning Curves}
\label{fig:ast_lc1}
\end{center}
\end{figure}

 A typical trajectory is shown below in Figure \ref{fig:ast_traj}, where the lower left hand subplot shows the value function's prediction of the value of the state at time $t$ . Learning curves are given in Figure \ref{fig:ast_lc1}. Test results are given in Table \ref{tab:exp1_perf}. We see that the DR/DV guidance law fails in this task, with occasional catastrophic position errors (these typically occur after less than 1000 episodes of testing). All of the RL derived policies have acceptable performance, but we see an increase in fuel efficiency as the number of recurrent steps in the forward pass is increased. 

\begin{figure}[htbp]
\begin{center}
\includegraphics[width=.9\linewidth]{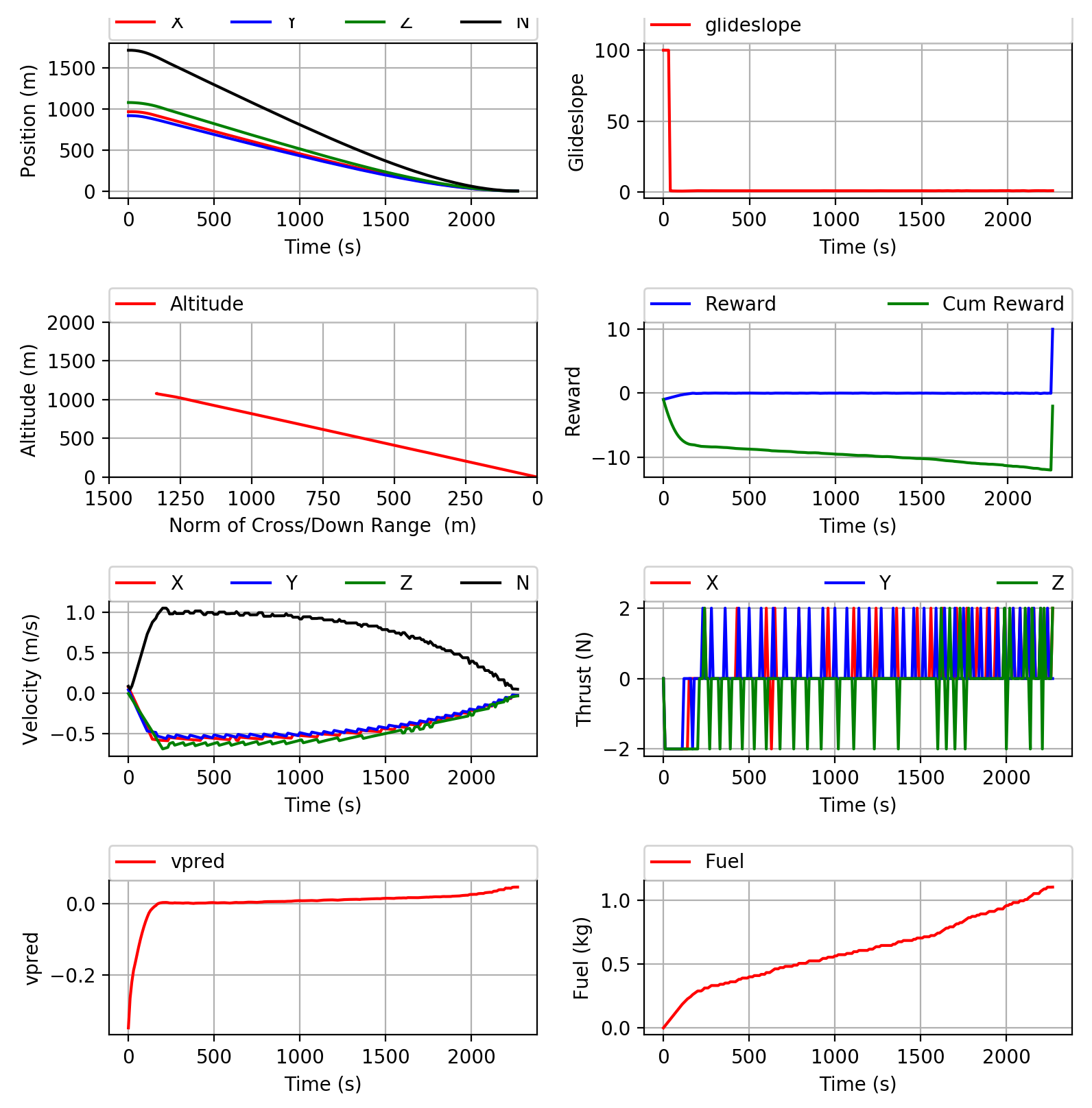}
\caption{Experiment 1: Typical Trajectory for Asteroid Landing}
\label{fig:ast_traj}
\end{center}
\end{figure}

\begin{table}[H]
	\fontsize{10}{10}\selectfont
    \caption{Experiment 1: Performance}
   \label{tab:exp1_perf}
        \centering 
   \newcolumntype{R}{>{\raggedleft\arraybackslash}p{0.8cm}}
   \begin{tabular}{l | R | R | R | R | R | R | R | R | R} 
      \hline 
      
       & \multicolumn{3}{c}{Terminal Position (m)}\vline & \multicolumn{3}{c}{Terminal Velocity (cm/s)}\vline & \multicolumn{3}{c}{Fuel (kg)}\\
       \hline
       & $\mu$ & $\sigma$ & max & $\mu$ & $\sigma$ & max & $\mu$ & $\sigma$ & max\\
       \hline
      DR/DV     & 1.9 & 54.6 & 1811 & 54 & 23 & 47 & 1.66 & 0.38 & 4.99\\
      MLP       & 0.2 & 0.1 & 0.9 & 4.1 & 1.4 & 9.7 & 1.62 & 0.92 & 3.32\\
      RNN 1 step & 0.2 & 0.1 & 0.7 & 3.8 & 1.3 & 8.7 & 1.48  & 0.73  & 3.36\\
      RNN 20 steps & 0.2 & 0.1 & 1.7  & 4.0 & 1.3 & 11.0 & 1.42 & 0.38  & 3.63 \\
      RNN 60 steps  & 0.3 & 0.2 & 1.0 & 4.2 & 1.3 & 9.0 & 1.44 & 0.37 & 3.37\\
      RNN 120 steps & 0.3 & 0.2 & 0.9 & 4.1 & 1.3 & 7.9 & 1.43 & 0.35 & 2.85 \\
   \end{tabular}
\end{table}

\subsection{Experiment2: Mars Landing using Radar Altimeter Observations}

\begin{figure}[htbp]
\begin{center}
\includegraphics[width=.9\linewidth]{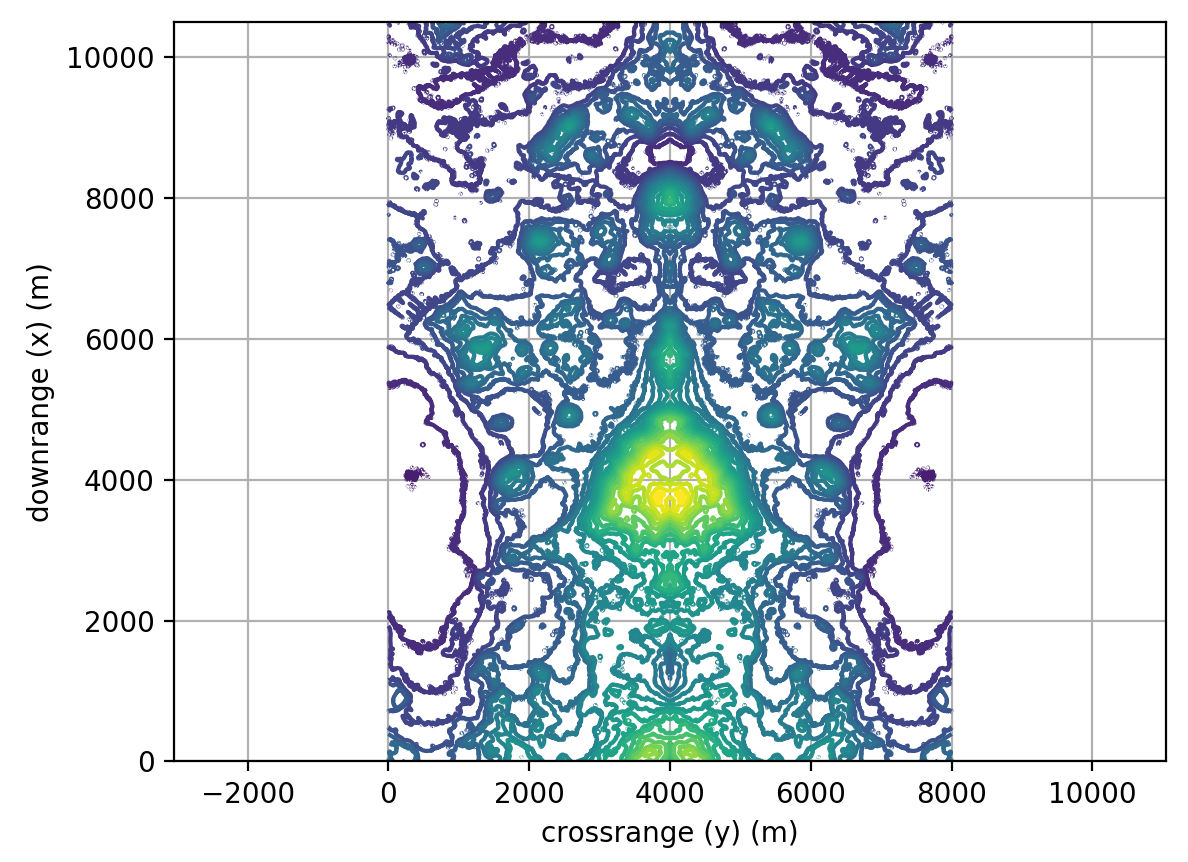}
\caption{Experiment 2: Digital Terrain Map}
\label{fig:dtm}
\end{center}
\end{figure}

In this task, the observations are simulated Doppler altimeter readings from the lander using a digital terrain map (DTM) of the Mars surface in the vicinity of Uzbois Valis. Since the simulated beams can cover a wide area of terrain, we doubled the map size by reflecting the map and joining it with original map, as shown in Figure \ref{fig:dtm}. Note that the agent does not have access to the DTM, but will learn how to successfully complete a maneuver using rewards received from the environment. Although these observations are a function of both lander position and lander velocity, the observations do not satisfy the Markov property as there are multiple ground truth positions and velocities that could correspond to a given observation, making the optimal action a function of the history of past altimeter readings. For the simulated altimeter readings, the agent's state in the target centered reference frame is transformed to the DTM frame, which has a target location of [4000, 4000, 400] meters.

In order to simulate altimeter readings fast enough to allow optimization to complete in a reasonable time, we had to create a fast altimeter model. The model uses a stack of planes with surface normals in the z (elevation direction) that spans the elevations in the DTM.  Each of the four radar beams has an associated direction vector, which, in conjunction with the spacecraft's position, can quickly be used to find the intersection of the beam and the planes.  The intersection x,y indices are used to index the DTM, and the plane intersection with a z value closest to that of the indexed DTM elevation is used to determine the distance between the lander and the terrain feature hit by the beam.  This is extremely fast (about 1000X faster than the ray-casting approach we used in Reference (\citenum{gaudet2014navigation}), but is error prone at lower elevations as sometimes the closest distance between DTM elevation and associated plane intersect z component is the far side of a terrain feature. Rather than call this a bug, we use it to showcase the ability of a recurrent policy to get remarkably close to a good landing, given the large errors.  The reduction in accuracy at lower elevations is apparent in Table \ref{tab:mm_acc}. The accuracy was estimated by choosing 10,000 random DTM locations and casting a ray to a random position at the designated elevation. The length of this ray is the ground truth altimeter reading.  We then checked what are measurement model returned from that lander position, with the error being the difference.  Note that the DTM elevations range from 0 to 380m.  In this scenario, the lander target's a landing position 50m above the top of a hill at 350m elevation.

\begin{figure}[htbp]
\begin{center}
\includegraphics[width=.9\linewidth]{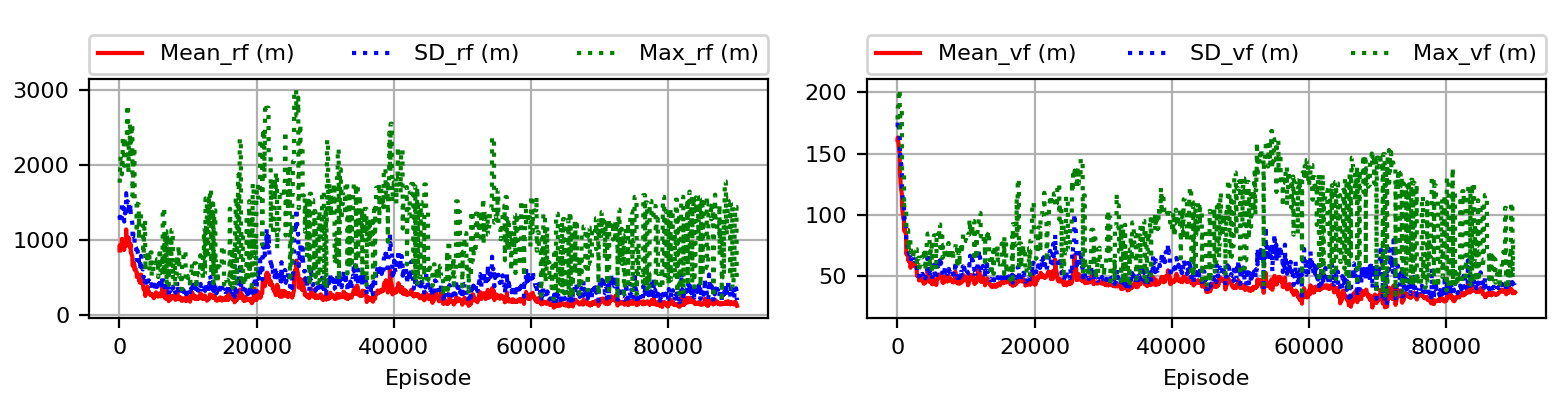}
\caption{Experiment 2: Learning Curves for 1 step RNN}
\label{fig:exp2_lc-1step}
\end{center}
\end{figure}

\begin{figure}[htbp]
\begin{center}
\includegraphics[width=.9\linewidth]{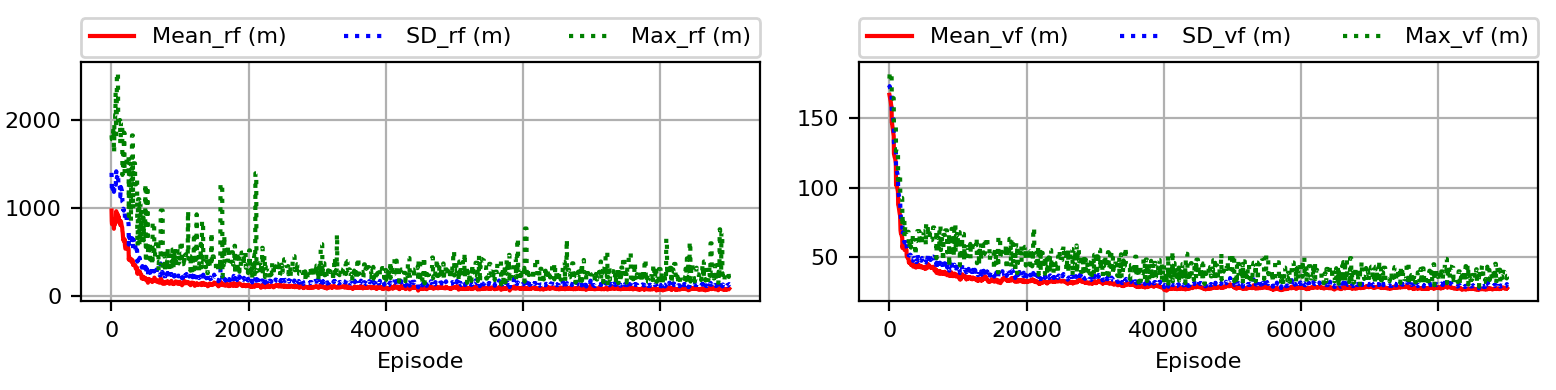}
\caption{Experiment 2: Learning Curves for 120 step RNN}
\label{fig:exp2_lc}
\end{center}
\end{figure}

The altimeter beams are modeled as having equal offset angles ($\pi/8$ radians) from a direction vector that points in a direction that is averaged between the lander's velocity vector and straight down. We thought this a reasonable assumption as we are modeling the lander in 3-DOF. We see from Table \ref{tab:dtm_perf1} that although you would not want to entrust an expensive rover to this integrated guidance and navigation algorithm, the performance is remarkably good given the altimeter inaccuracy at lower elevations. Learning curves for the 1-step and 20-step RNN's are shown in Figures \ref{fig:exp2_lc-1step} and \ref{fig:exp2_lc}, which plots statistics for terminal  position (r\_f) and terminal velocity (v\_f) as a function of episode, with the statistics calculated over the 30 episodes used to generate rollouts for updating the policy and value function.  We see from the learning curves that the amount of steps we unroll the recurrent network  in the forward pass has a large impact on optimization performance, and that for the 120 step case, the optimization initially makes good progress, but then stalls, likely due to the highly inaccurate altimeter readings at lower altitudes.

\begin{table}[htbp]
	\fontsize{10}{10}\selectfont
    \caption{Experiment2: Altimeter Error as function of lander elevation (m)}
   \label{tab:mm_acc}
        \centering 
   \newcolumntype{R}{>{\raggedleft\arraybackslash}p{1.8cm}}
   \begin{tabular}{l | R | R | R | R } 
       & mean (m) & std (m) & max (m) & miss \% \\
       \hline
       100 & 2600  & 3200 & 10000 & 61 \\
       400 & 513 & 1500 & 8800 & 22 \\
       500 & 122 & 528 & 4467 & 12 \\
       600 & 25 & 201 & 2545 & 6 \\
       700 & 8 & 92 & 1702 & 4 \\
       800 & 4 & 60 & 1300 & 2 \\
     
      \hline
    
   \end{tabular}
\end{table}

\begin{table}[htbp]
	\fontsize{10}{10}\selectfont
    \caption{Experiment 2: Performance}
   \label{tab:dtm_perf1}
        \centering 
   \newcolumntype{R}{>{\raggedleft\arraybackslash}p{0.80cm}}
   \begin{tabular}{l | R | R | R | R | R | R | R | R | R | R} 
      \hline 
      
       & \multicolumn{3}{c}{Terminal Position (m)}\vline & \multicolumn{3}{c}{Terminal Velocity (m/s)}\vline & \multicolumn{3}{c}{Glideslope} \vline & Fuel (kg) \\
       \hline
       & $\mu$ & $\sigma$ & max & $\mu$ & $\sigma$ & max & $\mu$ & $\sigma$ & min & $\mu$  \\
       \hline
      MLP       & 131 & 101 & 796 & 48  & 18 &  145 & 1.61  & 0.38 &  0.96 & 224 \\
      RNN 1 steps & 114 & 95 & 1359 & 37 & 4  & 67 & 3.57 & 3.87 &  0.49 & 228 \\
      RNN 20 steps & 78 & 41 & 390 & 26 & 1.7 & 39 & 3.54 & 6.05 & 0.49 & 234  \\
      RNN 120 steps & 72 & 40 & 349 & 28 & 2  & 44 & 3.34 & 3.78  & 0.49 & 233  \\
      RNN 200 steps & 59 & 42 & 288 &  23 & 4 & 40 & 2.55 & 2.62 & 0.70 & 242 \\
      
   \end{tabular}
\end{table}

In a variation on this experiment, we assume that the lander has the ability to point its radar altimeters such that the central direction vector remains fixed on the target location, and therefore the beams themselves bracket the target.  This functionality could be achieved with phased array radar, but a separate pointing policy would need to be learned that keeps the beams pointed in the required direction.  We also reduce the initial condition uncertainty to that shown in Table \ref{tab:IC_tp}.  Here we see that performance markedly improves.  We postulate one reason for the improved performance is that the altimeter beams remain in areas of high terrain diversity.  Indeed, when we repeat the experiment for a landing site further to the south (bottom of DTM), we find that performance degrades. Another factor could be that since the policy is focused on a small portion of the map, it does not "forget" the relationship between observations and advantages.

\begin{table}[htbp]
	\fontsize{10}{10}\selectfont
    \caption{Experiment 2: IC with Altimeter beams pointed at target}
   \label{tab:IC_tp}
        \centering 
   \newcolumntype{R}{>{\raggedleft\arraybackslash}p{1.8cm}}
   \begin{tabular}{l | R | R | R | R } 
      \hline 
       & \multicolumn{2}{c}{Velocity}\vline & \multicolumn{2}{c}{Position}\\
       \hline
       & min (m/s) & max (m/s) & min (m) & max (m) \\
       \hline
      Downrange      & -30 & -10 & 0 & 1000\\
      Crossrange       & -30  & 30 & -500 & 500 \\
      Elevation     & -50 & -40 & 1000 & 1000 \\
   \end{tabular}
\end{table}

\begin{table}[H]
	\fontsize{10}{10}\selectfont
    \caption{Experiment 2: Performance with target pointing}
   \label{tab:dtm_perf2}
        \centering 
   \newcolumntype{R}{>{\raggedleft\arraybackslash}p{0.8cm}}
   \begin{tabular}{l | R | R | R | R | R | R | R | R | R | R} 
      \hline 
      
       & \multicolumn{3}{c}{Terminal Position (m)}\vline & \multicolumn{3}{c}{Terminal Velocity (m/s)}\vline & \multicolumn{3}{c}{Glideslope} \vline & Fuel (kg) \\
       \hline
       & $\mu$ & $\sigma$ & max & $\mu$ & $\sigma$ & max & $\mu$ & $\sigma$ & min & $\mu$  \\
       \hline
      MLP       & 1.4  & 2.9 & 179.2  & 4.92 & 2.29 & 84.16 & 9.33 &  13.83 &  0.41 &  211\\
      RNN 1 step   & 3.3  & 1.9 & 108.6 &  5.75  & 1.38 & 69.80  & 6.42 &  6.42 & 0.30 & 220\\
      RNN 20 steps   &  0.3 & 1.2 &  116.0 & 1.61 & 0.60 & 64.84 & 11.22 & 15.73 & 0.98 & 219 \\
      RNN 60 steps   & 0.4  & 1.5 & 111.5 & 2.00 &  0.92 & 53.73 & 10.74 &  16.42 & 0.82 & 221\\
      RNN 120 steps & 0.6 & 1.6 & 139.8 & 2.23 & 0.94 & 57.47 & 8.32 & 14.67 & 0.74 & 223\\
   \end{tabular}
\end{table}

Taking into account the number of large outliers, it is probably best to focus on the average performance when comparing the network architectures. Fuel usage is really not a good measure of performance in this experiment, as it decreases with increased landing velocity.  The general trend is that performance increases as we increase the number of steps we unroll the recurrent layer for the forward pass. This implies that the temporal dependencies for this task probably span a significant fraction of a single episode.

\subsection{Experiment 3: Mars Landing with Engine Failure}

 To test the ability of the recurrent policy to deal with actuator failure, we increase the Mars lander's maximum thrust to 24000N. In a 6-DOF environment, each engine would be replaced by two engines with half the thrust, with reduced thrust occurring when one engine in a pair fails. At the start of each episode, we simulate an engine failure in 3-DOF by randomly choosing to limit the available downrange or crossrange thrust by a factor of 2, and limit the vertical (elevation) thrust by a factor of 1.5. Some episodes occur with no failure; we use a failure probability of 0.5. A real engine would hopefully be more reliable, but we want to optimize with each failure mode occurring often. The goal is to demonstrate a level of adaptability that would not be possible without an integrated and adaptive guidance and control system. We see in Table \ref{tab:actuator_failure} that the DR/DV policy fails catastrophically, whereas the MLP policy comes close to a safe landing.  The 1-step recurrent policy has performance close to that of the MLP policy, but performance improves to something consistent with a safe landing for the 20-step and 60-step recurrent policies.  The 20-step and 60-step policies also have improved fuel efficiency. 

\begin{table}[htbp]
	\fontsize{10}{10}\selectfont
    \caption{Experiment 3: Performance}
   \label{tab:actuator_failure}
        \centering 
   \newcolumntype{R}{>{\raggedleft\arraybackslash}p{0.8cm}}
   \begin{tabular}{l | R | R | R | R | R | R | R | R | R | R} 
      \hline 
      
       & \multicolumn{3}{c}{Terminal Position (m)}\vline & \multicolumn{3}{c}{Terminal Velocity (m/s)}\vline & \multicolumn{3}{c}{Glideslope} \vline & Fuel (kg) \\
       \hline
       & $\mu$ & $\sigma$ & max & $\mu$ & $\sigma$ & max & $\mu$ & $\sigma$ & min & $\mu$  \\
       \hline
      DR/DV     & 123 & 186 & 1061 & 20.07 & 19.06 & 59.29 & 2.16 & 1.04  & 0.96 & 213  \\
      MLP       & 0.7 & 0.2 & 1.8 &  0.99 & 0.62 & 4.67 & 13.46 & 5.35  & 4.93 & 302  \\
      RNN 1 steps & 0.9 & 0.2 & 1.7 & 0.95 & 0.44 & 4.76 & 14.23 & 3.44 & 6.29 & 298 \\
      RNN 20 step & 0.3 & 0.1 & 0.9 & 0.99 & 0.06 & 1.15  & 49.99  & 82.14 & 11.36 & 295 \\
      RNN 60 steps & 0.3 & 0.2 & 1.1 & 1.00 & 0.06 & 1.16 & 24.98 & 12.70 & 9.45 & 295 \\
   \end{tabular}
\end{table}

\subsection{Experiment 4: Mars Landing with High Mass Variation}

Here we divide the lander engine's specific impulse by a factor of 6, which increases fuel consumption to around 1200kg on average, with a peak of 1600kg. This complicates the guidance problem in that the mass varies by a significant fraction of the lander's initial mass during the descent, and we do not give the agent access to the actual mass during the descent. Although we are using a Mars landing environment for this task, the large variability in mass would be more similar to the problem encountered in an EKV interception of an ICBM warhead, where there is a high ratio of wet mass to dry mass.

We see in Table \ref{tab:Isp} that the DR/DV policy has a rather large maximum position error, and an unsafe terminal velocity.  The MLP policy and 1-step recurrent policy give improved performance, but still result in an unsafe landing. The 20-step recurrent policy achieves a good landing, which is slightly improved on by the 60-step recurrent policy.

\begin{table}[H]
	\fontsize{10}{10}\selectfont
    \caption{Experiment 4: Performance}
   \label{tab:Isp}
        \centering 
   \newcolumntype{R}{>{\raggedleft\arraybackslash}p{0.8cm}}
   \begin{tabular}{l | R | R | R | R | R | R | R | R | R | R} 
      \hline 
      
       & \multicolumn{3}{c}{Terminal Position (m)}\vline & \multicolumn{3}{c}{Terminal Velocity (m/s)}\vline & \multicolumn{3}{c}{Glideslope} \vline & Fuel (kg)\\
       \hline
       & $\mu$ & $\sigma$ & max & $\mu$ & $\sigma$ & max & $\mu$ & $\sigma$ & min & $\mu$ \\
       \hline
      DR/DV & 0.4 & 1.5 & 19.6 & 0.63 & 9.56 & 5.39 & 36.89 & 180.6 & 3.12 & 1362   \\
      MLP  & 0.7 & 0.2 & 3.7 & 0.92 & 0.26 & 5.25 & 10.12 & 2.09  &  0.63 & 1235   \\
      RNN 1 step & 0.4 & 0.1 & 0.8 & 0.98 & 0.42 & 6.48 & 20.05 & 4.20 & 7.71 & 1229 \\
      RNN 20 steps  & 0.6 & 0.1 & 1.0 & 1.17  & 0.07 & 1.36 & 22.64 & 7.18 & 12.62 & 1224  \\
      RNN 60 steps & 0.6 & 0.2 & 1.1 & 1.06 & 0.05 & 1.21 & 20.18 & 7.94 & 8.88 & 1227\\
   \end{tabular}
\end{table}

\section{Conclusion}

We have applied reinforcement meta learning with recurrent policy and value function to four difficult tasks: An asteroid landing with unknown and highly variable environmental parameters, a Mars landing using noisy radar altimeter readings for observations, a Mars landing with engine actuator failure, and a Mars landing with high mass variability during the descent. In each of these tasks, we find the best performance is achieved using a recurrent policy, although for some tasks the MLP policy comes close. The DR/DV policy had the worst performance over all four tasks. The take away is that the ability to optimize using parameter uncertainty leads  to robust policies, and the ability of a recurrent policy to adapt in real time to dynamic environments makes it the best performing option in environments with unknown or highly variable dynamics. Future work will explore the adaptability of recurrent policies in more realistic 6-DOF environments. We will also explore different sensor models for observations such as simulated camera images, and flash LIDAR for asteroid close proximity missions.

\bibliographystyle{AAS_publication}   
\bibliography{references}   

\begin{thebibliography}{10}

\bibitem{guang2018attitude}
Z.~Guang, Z.~Heming, and B.~Liang, ``Attitude Dynamics of Spacecraft with
  Time-Varying Inertia During On-Orbit Refueling,''  {\em Journal of Guidance,
  Control, and Dynamics}, 2018, pp.~1--11.

\bibitem{yu2017preparing}
W.~Yu, J.~Tan, C.~K. Liu, and G.~Turk, ``Preparing for the unknown: Learning a
  universal policy with online system identification,''  {\em arXiv preprint
  arXiv:1702.02453}, 2017.

\bibitem{peng2017sim}
X.~B. Peng, M.~Andrychowicz, W.~Zaremba, and P.~Abbeel, ``Sim-to-real transfer
  of robotic control with dynamics randomization,''  {\em arXiv preprint
  arXiv:1710.06537}, 2017.

\bibitem{gaudet2018deep}
B.~Gaudet, R.~Linares, and R.~Furfaro, ``Deep Reinforcement Learning for Six
  Degree-of-Freedom Planetary Powered Descent and Landing,''  {\em arXiv
  preprint arXiv:1810.08719}, 2018.

\bibitem{d1997optimal}
C.~D'Souza and C.~D'Souza, ``An optimal guidance law for planetary landing,''
  {\em Guidance, Navigation, and Control Conference}, 1997, p.~3709.

\bibitem{mnih2015human}
V.~Mnih, K.~Kavukcuoglu, D.~Silver, A.~A. Rusu, J.~Veness, M.~G. Bellemare,
  A.~Graves, M.~Riedmiller, A.~K. Fidjeland, G.~Ostrovski, {\em et~al.},
  ``Human-level control through deep reinforcement learning,''  {\em Nature},
  Vol.~518, No.~7540, 2015, p.~529.

\bibitem{baccouche2011sequential}
M.~Baccouche, F.~Mamalet, C.~Wolf, C.~Garcia, and A.~Baskurt, ``Sequential deep
  learning for human action recognition,''  {\em International Workshop on
  Human Behavior Understanding}, Springer, 2011, pp.~29--39.

\bibitem{williams1992simple}
R.~J. Williams, ``Simple statistical gradient-following algorithms for
  connectionist reinforcement learning,''  {\em Machine learning}, Vol.~8,
  No.~3-4, 1992, pp.~229--256.

\bibitem{schulman2017proximal}
J.~Schulman, F.~Wolski, P.~Dhariwal, A.~Radford, and O.~Klimov, ``Proximal
  policy optimization algorithms,''  {\em arXiv preprint arXiv:1707.06347},
  2017.

\bibitem{schulman2015trust}
J.~Schulman, S.~Levine, P.~Abbeel, M.~Jordan, and P.~Moritz, ``Trust region
  policy optimization,''  {\em International Conference on Machine Learning},
  2015, pp.~1889--1897.

\bibitem{sorensen1982newton}
D.~C. Sorensen, ``Newton’s method with a model trust region modification,''
  {\em SIAM Journal on Numerical Analysis}, Vol.~19, No.~2, 1982, pp.~409--426.

\bibitem{kullback1951information}
S.~Kullback and R.~A. Leibler, ``On information and sufficiency,''  {\em The
  annals of mathematical statistics}, Vol.~22, No.~1, 1951, pp.~79--86.

\bibitem{chung2015gated}
J.~Chung, C.~Gulcehre, K.~Cho, and Y.~Bengio, ``Gated feedback recurrent neural
  networks,''  {\em International Conference on Machine Learning}, 2015,
  pp.~2067--2075.

\bibitem{ng2000algorithms}
A.~Y. Ng, S.~J. Russell, {\em et~al.}, ``Algorithms for inverse reinforcement
  learning.,''  {\em Icml}, 2000, pp.~663--670.

\bibitem{ng2003shaping}
A.~Y. Ng, {\em Shaping and policy search in reinforcement learning}.
\newblock PhD thesis, University of California, Berkeley, 2003.

\bibitem{schulman2015high}
J.~Schulman, P.~Moritz, S.~Levine, M.~Jordan, and P.~Abbeel, ``High-dimensional
  continuous control using generalized advantage estimation,''  {\em arXiv
  preprint arXiv:1506.02438}, 2015.

\bibitem{udrea2012sensitivity}
B.~Udrea, P.~Patel, and P.~Anderson, ``Sensitivity Analysis of the Touchdown
  Footprint at (101955) 1999 RQ36,''  {\em Proceedings of the 22nd AAS/AIAA
  Spaceflight Mechanics Conference}, Vol.~143, 2012.

\bibitem{gaudet2014navigation}
B.~Gaudet and R.~Furfaro, ``A navigation scheme for pinpoint mars landing using
  radar altimetry, a digital terrain model, and a particle filter,''  {\em 2013
  AAS/AIAA Astrodynamics Specialist Conference, Astrodynamics 2013}, Univelt
  Inc., 2014.

\end{thebibliography}

\end{document}